\documentclass{IEEEcsmag}

\usepackage[colorlinks,urlcolor=blue,linkcolor=blue,citecolor=blue]{hyperref}

\usepackage{amsmath}
\usepackage{upmath}
\usepackage{comment}
\usepackage{tabularx}
\usepackage{caption}
\usepackage{subcaption}
\usepackage{booktabs}
\usepackage{float}

\jvol{XX}
\jnum{XX}
\paper{XX}
\jmonth{}
\jname{IEEE Security \& Privacy}
\pubyear{2022}

\begin{document}


\title{Phishing Detection Leveraging Machine Learning and Deep Learning: A Review}

\author{Dinil Mon Divakaran}
\affil{Trustwave}

\author{Adam Oest}
\affil{PayPal, Inc.}


\begin{abstract}
Phishing attacks trick victims into disclosing sensitive information. To counter rapidly evolving attacks, we must explore machine learning and deep learning models leveraging large-scale data. We discuss models built on different kinds of data, along with their advantages and disadvantages, and present multiple deployment options to detect phishing attacks.

\end{abstract}
\maketitle


\begin{figure}
\centerline{\fbox{\includegraphics[scale=0.16]{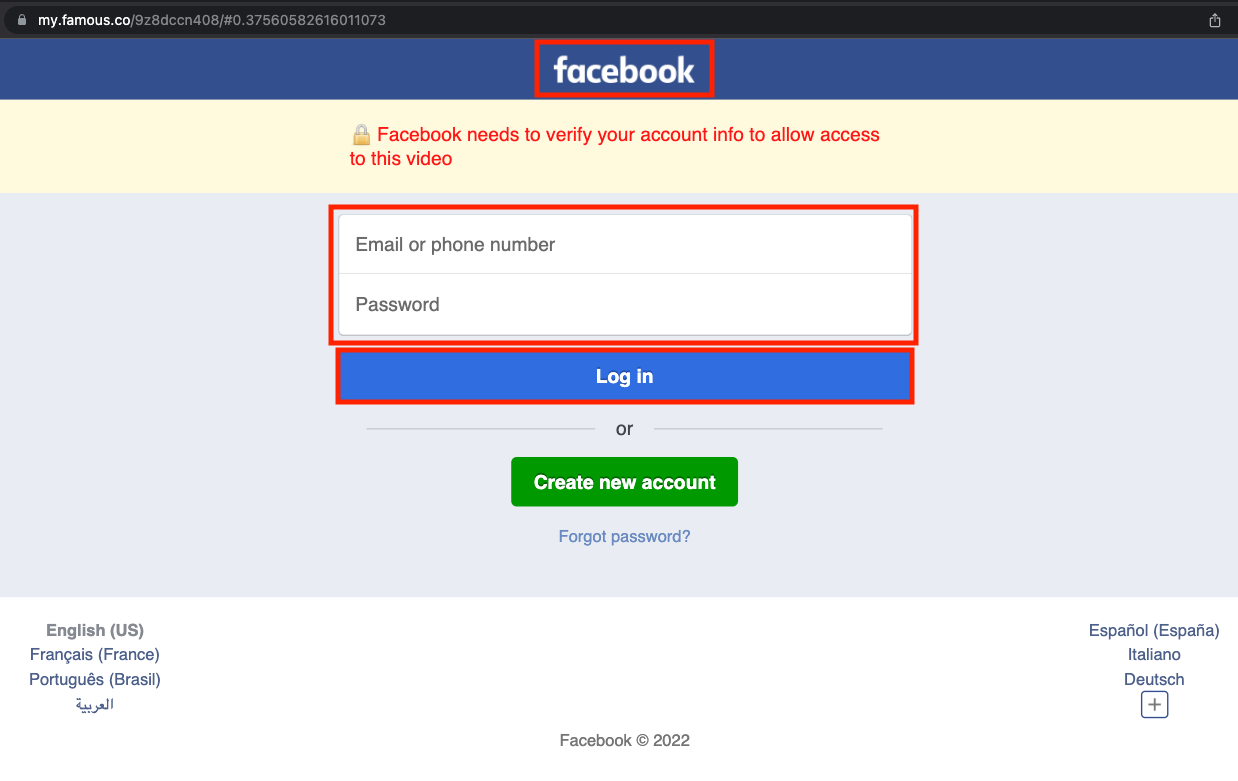}}}
\caption{A phishing page targeting Facebook (screenshot taken in March 2022). Red boxes highlight important phishing characteristics---logo, input form and login button---designed to deceive a user.}
\label{fig:phishing-example-FB}
\end{figure}

\begin{figure*}[ht!]
\centerline{\includegraphics[scale=0.15]{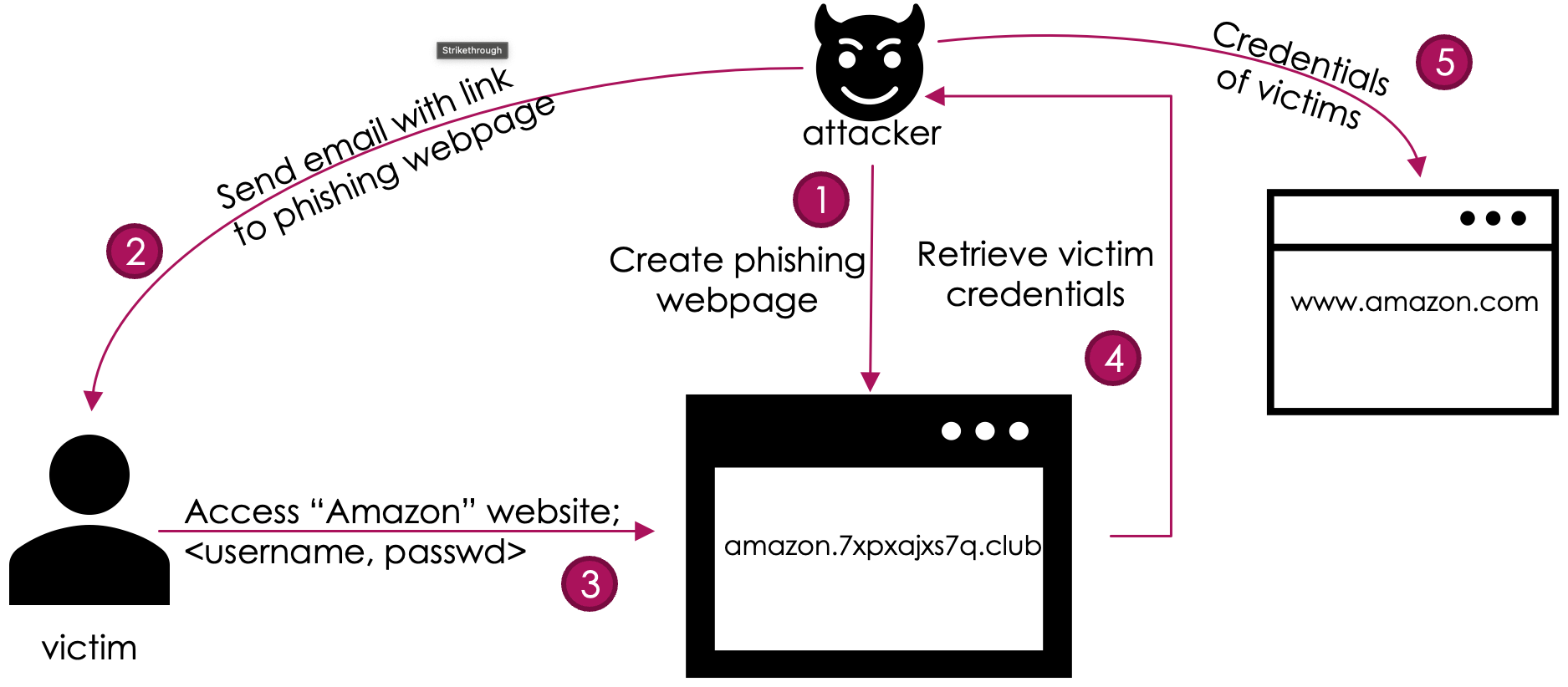}}
\caption{Illustration of phishing. 1) An attacker deploys a phishing webpage mimicking a well-known legitimate target, e.g., Amazon. 2) The attacker sends the corresponding link to numerous potential victims via emails, social networks, etc. 3) A user who gets deceived accesses the  phishing webpage and provides sensitive information. 4) The user credentials are retrieved by the attacker, 5) to access the targeted website to commit fraud. }
\label{fig:phishing-illus}
\end{figure*}

\chapterinitial{P}hishing attacks attempt to deceive users with the goal of stealing sensitive information such as user credentials (username and password), bank account details, and even entire identities. To achieve this, 
an attacker deploys a malicious website that resembles the legitimate website of a well-known target brand, e.g., {\bf Figure~\ref{fig:phishing-example-FB}} shows a real phishing page targeting Facebook. Subsequently, the attacker sends a link to the phishing website to potential victims and leverages social engineering techniques to lure those victims into disclosing confidential information.  {\bf Figure~\ref{fig:phishing-illus}} illustrates the general process, which can be quite complicated in practice. A recent phishing campaign used a compromised email account to send victim users a malicious attachment containing obfuscated code.  This code, in turn, retrieved dynamic scripts from a hosting server that ultimately generated HTML to render a phishing page targeting Microsoft Office 365 users (for a detailed description, see~\cite{trustwave-phishing-campaign-2021}). Indeed, phishing attacks have evolved to increasingly use shortened URLs, redirection links, the HTTPS protocol, server- and client-side cloaking techniques (displaying benign pages to anti-phishing crawlers), 
etc., to evade detection~\cite{zhang2021crawlphish,phishing-life-cycle-2020}.
To add to these, today there exist software tool kits, or {\em phishing kits}, in dark markets, that automate the majority of the aforementioned process.  
Thus, it is no surprise that phishing continues to be one of the top cyber attacks today.  Furthermore, the COVID-19 pandemic and the prevalence of remote work have given rise to new social engineering and victimization opportunities for attackers~\cite{phishing-pandemic-2020}.

\subsection{Anti-Phishing Blacklists}

A well-known technique for protecting users from accessing phishing webpages is to use an anti-phishing blacklist. Webpage URLs accessed from a modern web browser (e.g., Google Chrome, Mozilla Firefox, or Microsoft Edge) are automatically checked against a list of known phishing URLs; and access to a matched URL is warned/blocked.  The URLs that ultimately end up on anti-phishing blacklists are collected, crawled, and analyzed by anti-phishing entities (e.g., Google Safe Browsing, the Anti-Phishing Working Group, or PhishTank), each of which leverages a variety of threat intelligence sources and cross-organizational partnerships.

Blacklists are efficient and scale well with the number of phishing webpages in the wild. A URL lookup can easily be done in real-time, using the user's computer resources, and such lookups also preserve user privacy.  However, anti-phishing blacklists are not themselves capable of detecting new and unseen phishing pages: they rely on crawlers that must analyze the webpage content beforehand.  Such crawlers are, in turn, susceptible to evasion attacks~\cite{oest2020phishtime} that make phishing webpages appear benign whenever they are accessed by a crawler. Blacklist-based approaches also inherently struggle when attackers use redirection links as lures, as these benign-looking links differ from the URL of the corresponding phishing webpage.  
These limitations of large-scale anti-phishing systems have motivated researchers to explore learning algorithms to develop phishing detection solutions capable of running in multiple different contexts, that can keep up with the evasion efforts of attackers.

\section{Supervised Machine Learning: a brief overview}

\begin{figure}
     \centering{
     \begin{subfigure}[b]{0.48\textwidth}
         \centering
         \includegraphics[width=\textwidth]{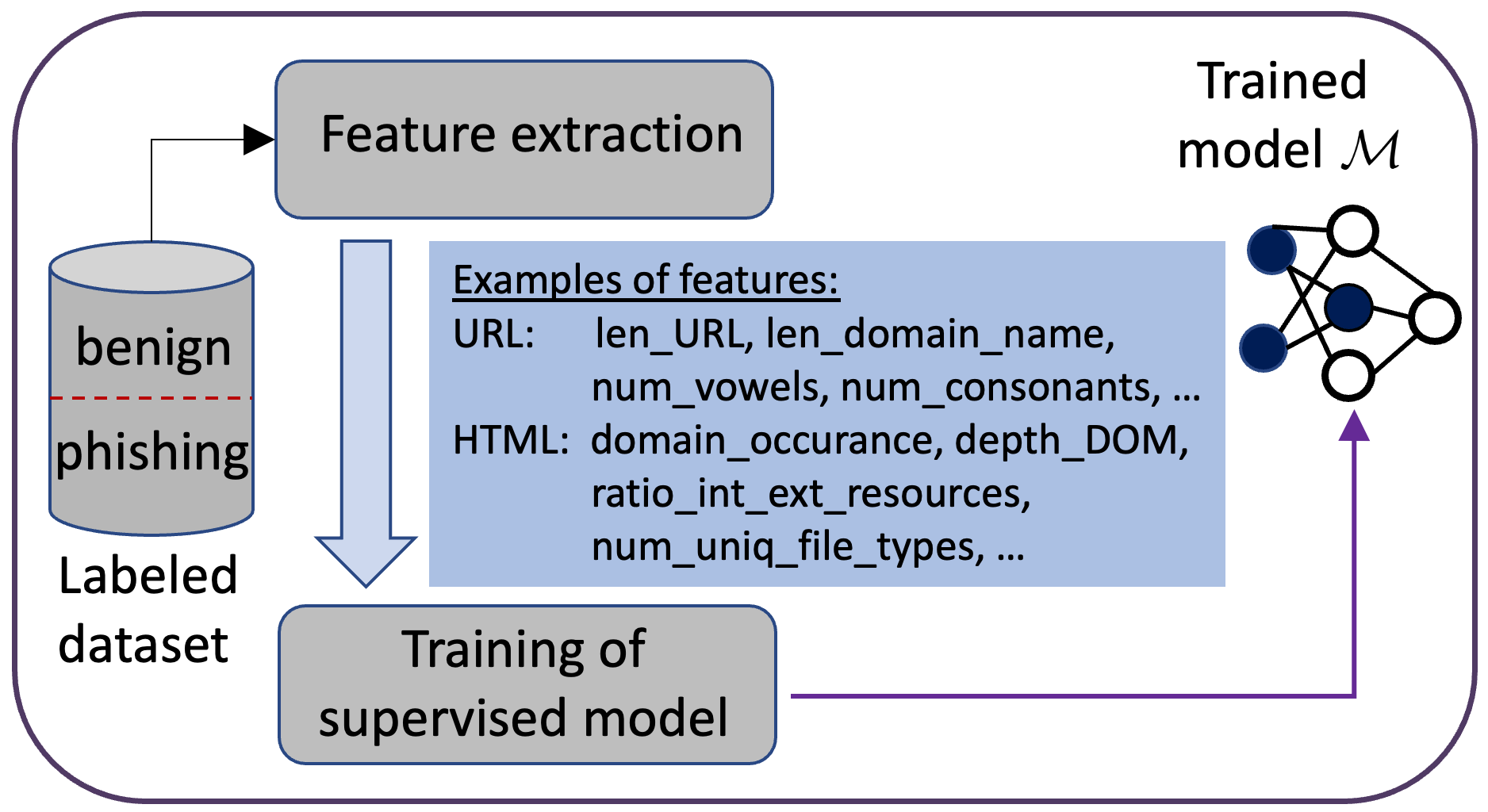}
         \caption{Training phase}
         \label{fig:ML-training}
     \end{subfigure}
     \hfill
     \begin{subfigure}[b]{0.39\textwidth}
         \centering
         \includegraphics[width=\textwidth]{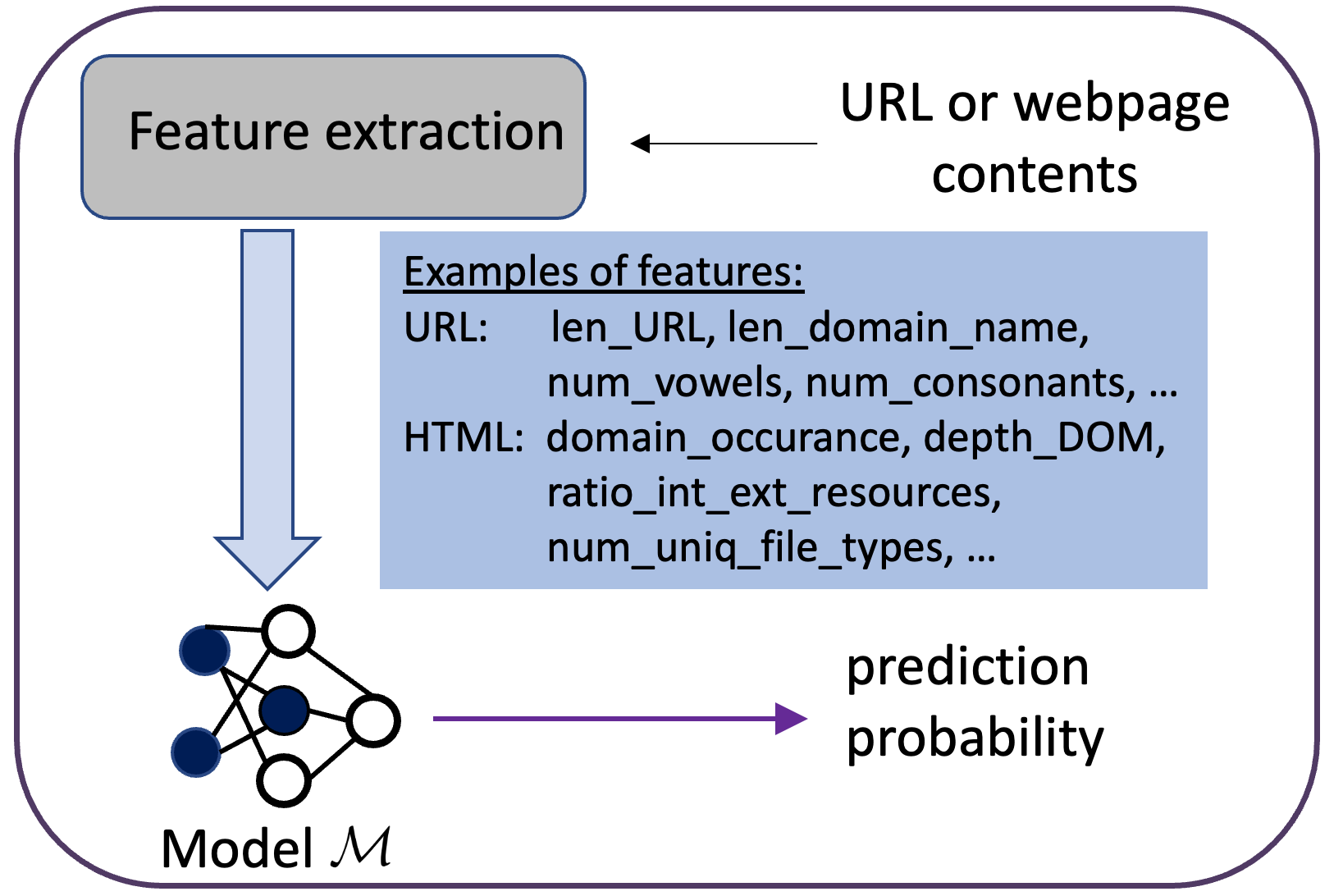}
         \caption{Inference phase}
         \label{fig:ML-testing}
     \end{subfigure}
\caption{Pipeline for building supervised phishing detection model. (a) Model $\mathcal M$ is trained offline, using labeled dataset, and (b) used for inference during online operation.}
\label{fig:ML-pipeline}}
\end{figure}

A supervised machine learning (ML) algorithm takes a large labeled dataset as input to train a classification model that subsequently classifies an input data point into a given number of classes. {\bf Figure~\ref{fig:ML-pipeline}} presents an ML pipeline for developing supervised models that detect phishing attacks.  
In a phishing webpage detection problem, there are only two classes---{\em benign} and {\em phishing}; and hence, the trained models are binary classifiers. Each data point (e.g., a URL) in the input dataset is accompanied by a ground-truth label of {\em benign} or {\em phishing}, to help the model learn the discriminative characteristics of the two classes. Let $\mathcal M$ denote such a trained model. 
In operation (also called the inference phase), this trained model $\mathcal M$ is 
given a webpage URL, and it computes the probability $p$ that the webpage is a phishing webpage. 
Given a configuration threshold $\theta$, the input (URL) is classified as phishing if the prediction probability is greater than or equal to this threshold, i.e., if $p \ge \theta$; and benign otherwise.

\subsection{Performance metrics}

The accuracy of a model is dependent on a few factors, some of the important ones being, the model algorithm, the quality and quantity of data used for training, and the value of classification threshold used in operation. A phishing webpage correctly classified is considered a {\em True Positive} (TP) and a benign webpage correctly classified is a {\em True Negative} (TN). Conversely, a phishing webpage wrongly classified (as a benign page) is a {\em False Negative} (FN) and a benign webpage misclassified as phishing is a {\em False Positive} (FP).
The most important metrics for evaluating phishing detection models are
i)~the True-Positive Rate (TPR),  also called {\em Recall}, and ii)~the False-Positive Rate (FPR). 
The Recall of a model is defined as $\frac{\text{TP}}{\text{TP}+\text{FN}}$; and the FPR is defined as $\frac{\text{FP}}{\text{FP}+\text{TN}}$. 

The performance of a classification model is a trade-off between Recall and FPR; e.g., a 100\% Recall can be achieved by classifying every webpage as phishing (i.e., when $\theta = 0$), but that would also result in the highest FPR for that model. 
Therefore, a good evaluation strategy is to measure a model's Recall at varying values of FPR. An FPR of $10^{-1}$ means, on average 1~in~10 benign webpages visited by a user are misclassified as phishing. If the users are accordingly warned or blocked, then an FPR of $10^{-1}$ is clearly unacceptable due to the high number of disruptions caused to legitimate browsing. Therefore, in practical deployments,  phishing detection solutions need to evaluate Recall at FPR values of $10^{-3}$ and even lower~\cite{phishpedia-usenix-sec-2021, d-fence-2021}.

\section{Learning Algorithms for Phishing Detection}

Below we discuss different machine learning techniques for phishing detection. We distinguish the solutions based on the input they process (for training and prediction)---URLs, HTML contents, and webpage screenshots---as this leads to different deployment use cases. {\bf Table~\ref{tab:taxonomy}} provides a taxonomy of such detection techniques.

\subsection{Phishing URL Classification}

A phishing URL classifier trains a classification model on a dataset of benign and phishing URL strings along with the ground-truth labels (i.e., in a supervised way). 
Researchers often use top-ranking websites from Alexa (https://www.alexa.com/topsites) or Tranco (https://tranco-list.eu) to build a dataset of benign URLs. Whereas, the set of phishing URLs is usually obtained from PhishTank (https://phishtank.org/) and OpenPhish (https://openphish.com/). Once a model is trained, in operation, it is given a URL obtained from a user (e.g., from a user's browser or email, or from a middlebox such as a web-proxy) to make a prediction of whether it links to a phishing page or not. Note that, during both the training and the prediction phases, these models process only the URLs to extract useful information called {\em features}; they do not visit the given URLs. Thus, there is no additional cost of webpage visit, network latency, or  page-load time for a URL-based ML solution. A URL
can be readily processed, and a prediction can be made before the corresponding webpage is visited.

\begin{table*}[th]
\centering
\caption{Taxonomy of phishing detection solutions based on learning models \label{tab:taxonomy}}
{
\begin{tabular}{cp{1.6cm}p{3.6cm}p{3.7cm}p{3.7cm}} \toprule
\# & Data &  Model &  Advantage & Disadvantage \\ \midrule
1 & URL strings &   ML-based model with features engineered from URL strings  &  Fast, no network latency involved. & Limited information in URLs. Manual engineering of features.  \\  \hline

2 & URL strings &   DL-based, with URL strings provided as input to the neural network  &  Fast, no network latency involved. No feature engineering required. & Limited information in URLs. \\ \hline

3 & Webpage contents   & ML-based, with features engineered from HTML body (in addition to the URL) of the webpage & Contents provide rich information for models to learn and differentiate between benign and phishing webpages. & Network latency in obtaining the page contents, and expensive feature extraction for real-time detection. Susceptible to evasion techniques.\\ \hline
4 & Webpage screenshot   &  DL models to compare webpage screenshots and visual invariants such as logos & Not dependent on large labeled (benign and phishing) datasets. Instead, only a small reference list of top targeted websites needs to be maintained. & Cost of network latency to fully load a webpage and take a screenshot. Prone to adversarial ML attacks. \\
\bottomrule
\end{tabular}
}
\end{table*}

\subsubsection{ML-based Phishing URL Classifier}
\label{subsubsec:ML-URL}

Early classification solutions engineered {\em features} from URLs to train machine learning models ({Table~\ref{tab:taxonomy}}, row 1). The features try to capture the lexical properties of the URLs; they include, the length of the URL, the length of the domain part of the URL, the length of the path in the URL, the number of permitted non-alphanumeric characters (such as dot and hyphen) encoded in a URL, the number of vowels and consonants, etc. Besides, more complicated features are engineered by learning a vocabulary of words (or tokens), and representing the presence of each word in the URL (separately for different components of the URL) as a long binary vector~\cite{URL-based-ICML-2009}.
This results in long {\em feature vectors}, and affects the performance of some of the conventional ML models, such as SVM (support vector machine). There are also external features such as IP and domain reputation, domain registration information from WHOIS database, etc., that are useful for modeling. However, extraction of these external features requires additional network latency or maintenance of up-to-date large databases.

A drawback here is that, one has to {\em manually} define and engineer discriminative features; and the large body  of research works in this direction is evidence that defining all relevant high-quality features exhaustively is a challenging task. Furthermore, since the phishing attack ecosystem keeps evolving, the list of features needs to be continuously updated. It is likely that, at any point in time, the list of features used in a model is incomplete. 
Thus, more recent research works apply deep learning (DL) models to detect phishing URLs. 

\subsubsection{DL-based Phishing URL Classifier}
Convolutional Neural Network (CNN) is a state-of-the-art deep learning model that has been often used for visual analysis, while also being applied for text classification. For building a DL-based phishing URL classifier ({Table~\ref{tab:taxonomy}}, row 2), 
we first transform the input to a vector of numbers, referred to as {\em embedding}.  
Each character in a URL is represented by a vector of fixed dimension, and thus each URL is represented as a matrix, with each row representing a character. 
Since the matrix dimension is also fixed, the number of characters processed in a URL is pre-determined; i.e., the length of URLs processed is limited. A CNN model essentially performs convolution operations on the input at its different nodes, to learn the differentiating patterns between phishing and benign URLs in the labeled dataset; this model has been shown to be effective in detecting phishing URLs~\cite{URLnet-2018}. However, character-level models might not be able to capture relationship between characters, and more importantly words, that are far apart in a URL. 
This is where a sequence model, often used in language modeling, is useful. The LSTM (long short-term memory) architecture, a widely used sequence model, learns dependencies between characters in a long sequence (or string, as in a URL). The basic unit of an LSTM has multiple {\em gates} that allow it to learn (and forget) information from arbitrary points in the sequence. 
Lee et al.~\cite[Section 6.1]{d-fence-2021}
evaluated the performance of CNN, LSTM, as well as a combined architecture of CNN and LSTM  on a large dataset of URLs obtained from enterprise customers.
At a low FPR of $10^{-3}$, the combined CNN-LSTM model performed significantly better (achieving a Recall of $\approx76\%)$ than the independent models (that achieved $\approx 58\%$ Recall) in detecting phishing URLs. 
In this combined model, the output of CNN is fed to LSTM, and that of LSTM is used for prediction. The architecture consists of an embedding layer, followed by 256 Conv1D filters, a pooling layer (using MaxPooling), 512 LSTM units, followed by a Dropout layer, a Dense layer, and finally a single unit for classification (using sigmoid activation function).

\subsubsection{Limitations} 
The information available to train a classification model is limited to what is available in a URL. An attacker today has many options to decide on a domain name, the prominent part of a URL; and the rest of the URL (following the domain name), i.e., the path of a file under the website, is completely under the attacker's control. 
This makes it easy for an attacker to evade a URL-based detection model. Besides, shortened URLs present much less information for a model to make a good prediction. It is therefore not surprising that URLNet~\cite{URLnet-2018}, a URL-based deep learning model, incurs large number of false positives in a recent phishing discovery study conducted in the wild (Internet)~\cite[Section 6]{phishpedia-usenix-sec-2021}. One way to overcome this limitation is to develop models based on webpage contents and screenshots. We discuss them below.

\subsection{Phishing Content Classification}

The content of webpages offers a wealth of information that can be exploited to detect phishing attacks ({Table~\ref{tab:taxonomy}}, row 3). 
Since the goal of an attacker is to deceive users to provide their sensitive information, phishing webpages often have HTML forms and other discriminating characteristics. Therefore, the features that capture the presence of HTML forms, \texttt{<input>} tags, and sensitive keywords that prompt users to input username, password, credit card numbers, etc., are useful in building an HTML-based phishing classifier~\cite{canita+-2011}. There are also a number of other features useful for modeling; they include the length of the HTML \texttt{<form>}, \texttt{<style>}, \texttt{<script>} and comments, the number of words in text and title, the number of images and iframes, the presence of hidden content, the ratio of domain names in the HTML body, and the ratio of the number of external links to the number of links under the same domain.
In addition, features from URLs are used along with features engineered from the webpage contents. 
Once the features are defined, a supervised model, e.g., Random Forest, is trained using a set of labeled phishing and benign webpages (e.g., from feeds from Tranco and PhishTank); note, these webpages typically have to be crawled to create such a dataset.
Among the set of URL and HTML features considered, 
experiments show 
`the number of times the domain name (of the URL) appears in the HTML contents,'  `the number of unique subdomains (of the main domain in the URL) present in the HTML contents,' and `the number of unique directory paths for all files referenced in the HTML contents' are important ones~\cite{phishing-NDSS-MADWeb-2020}. Although the top features might differ between studies depending on the time and the source of data (in addition to other aspects such as the model used), the conclusion that HTML features tend to be more useful than URL features in the classification of phishing pages has been observed in another recent study~\cite{stack-model-phishing-2019}.

\subsubsection{Limitations} 
In comparison to a URL-based model, although the content-based model is more accurate, the process of feature extraction from webpage contents is expensive~\cite[Section 9.2]{canita+-2011}. The cost of feature extraction during model training is tolerable (since it is an offline process), but during inference, this cost affects the latency experienced by a user. Before allowing a user to access a  webpage, the model has to extract features from the webpage and make an inference. 
The HTML-based models are also susceptible to evasion attacks. For example, the three important features mentioned above can be manipulated by an attacker to evade a phishing classifier. As demonstrated in~\cite{phishing-NDSS-MADWeb-2020}, one way to counter such an attack (while maintaining detection capability) is to train the classifier with added noise in such a way that, the feature importance distribution becomes uniform instead of being skewed to a small set of features.
Besides carefully designing  HTML pages and contents, attackers also employ {\em cloaking techniques}~\cite{zhang2021crawlphish}. For example, the use of JavaScript for dynamic content rendering (often requiring user interaction) makes it challenging for automated phishing detection solutions to extract the content of malicious web pages. A recent phishing campaign targeting Office~365 users employed multiple JavaScript codes hosted on another server, to stack together a page mimicking Microsoft Office 365 login interface using Microsoft logos~\cite{trustwave-phishing-campaign-2021}. The page prompts users to provide their credentials to log in to their accounts.  To counter such evasion techniques, an interesting approach is to directly analyze the screenshots of webpages.

\subsection{Phishing Detection based on Webpage Screenshot}

Phishing websites are effective in deceiving users due to the visual similarity of a phishing webpage to a well-known website being impersonated. A phishing page targeting, say, Facebook users might try to imitate the look and style of Facebook website (see Figure~\ref{fig:phishing-example-FB}). To achieve visual similarities, the attackers use logos of the target brand/company (e.g., of Facebook's), the structure, style, and color of the legitimate website (\texttt{www.facebook.com}), the content of the target website, etc., while deploying a phishing webpage. This understanding opens up a different direction  of research for detecting phishing pages---that which looks for webpages impersonating well-known legitimate brands, {\em but} having domain names different from the targeted brands. For example, if a webpage looks visually similar to that of Facebook's  (\texttt{www.facebook.com}), but has a different domain (say, \texttt{my.famous.co/9z8dccn408}), then the webpage is very likely a phishing page.

\subsubsection{DL for Screenshot-based Phishing Detection} The tremendous success of deep learning in the domain of Computer Vision has spurred new solutions in screenshot-based phishing detection ({Table~\ref{tab:taxonomy}}, row 4). Siamese networks and its enhancements learn the similarity of the vector representations of given images. This is achieved by training the model in a supervised way, with  positive (similar) images and negative (dissimilar) images, via a loss function that decreases the dissimilarity of positive images (since they are supposed to look similar) and increases the dissimilarity of negative images. Put differently, the model is trained to differentiate between similar and dissimilar images. 
In VisualPhishNet~\cite{visualphishnet}, screenshots from a {\em reference list} of {\em protected} websites are used to train a Siamese model; subsequently, in operation, the trained model estimates the similarity between a given (potentially) suspicious webpage screenshot and all the screenshots in the reference list. A webpage that has high similarity with any website in the reference list while having a different domain is considered a phishing page. 

\begin{figure}
     \centering{
     \begin{subfigure}[b]{0.46\textwidth}
         \centering
         \includegraphics[width=\textwidth]{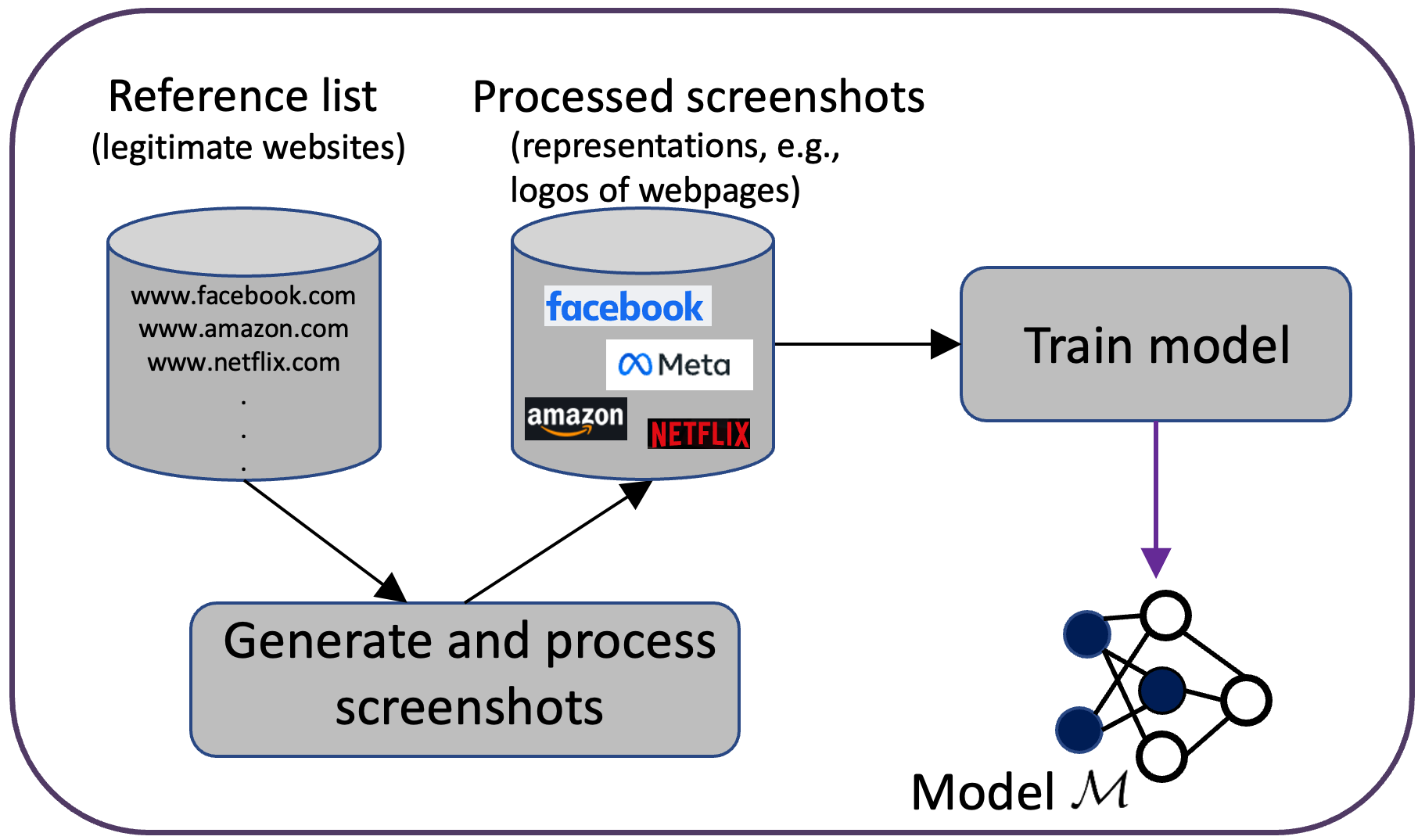}
         \caption{Training phase: model $\mathcal M$ trained using data generated from a reference list of legitimate websites}
         \label{fig:ref-training}
     \end{subfigure}
     \begin{subfigure}[b]{0.49\textwidth}
         \centering
         \includegraphics[width=\textwidth]{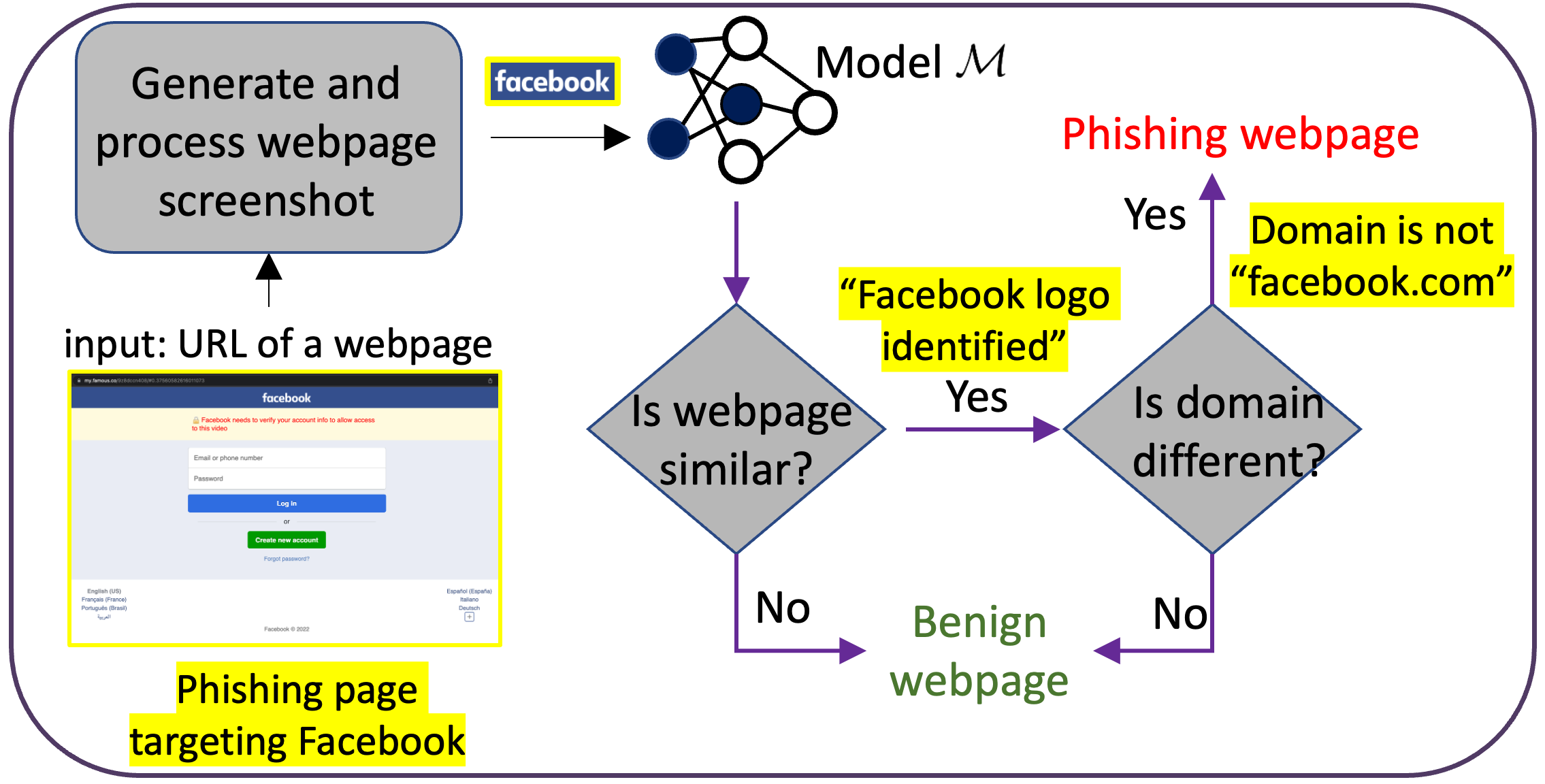}
         \caption{Inference phase: Trained model $\mathcal M$ used in operation. Highlights in yellow illustrate detection of an example  phishing page (given in Figure~\ref{fig:phishing-example-FB})}
         \label{fig:ref-testing}
     \end{subfigure}
\caption{Pipeline for screenshot-based phishing detection}
\label{fig:ref-pipeline}}
\end{figure}

Another idea is to infer if a given webpage uses logos of websites in the reference list of well-known companies and brands while having a different domain. The problem is broken into two sub-problems: one to detect and extract logos from a given webpage, and the other to match the extracted logo(s) with those in the reference list of websites. 
A recent proposal, Phishpedia~\cite{phishpedia-usenix-sec-2021}, uses an object detection model for the first task and a Siamese network additional the second task. Another work went further, modeling additional salient UI features such as input forms, buttons, text label, and block structures on screenshots~\cite{phishintention-usenix-sec-2022}. Both VisualPhishNet and Phishpedia demonstrate high accuracy in detecting phishing webpages, and they achieve this by analyzing the similarities with known images (screenshots and logos on screenshots), which in turn are learned from a reference list of websites. Thus the main dependency is only on a small list of websites that form the protected reference list. A reference list of top 100 commonly targeted legitimate websites covers around 95\% of all phishing attacks~\cite{phishpedia-usenix-sec-2021, phishintention-usenix-sec-2022}. Thus, VisualPhishNet and Phishpedia use a reference list of only a few hundred websites to achieve a high detection rate.  A general pipeline for such screenshot-based phishing detection solutions is depicted in {\bf Figure~\ref{fig:ref-pipeline}}. Different from URL-based and content-based approaches, screenshot-based solutions that use reference lists typically do not require a large labeled dataset of both benign and phishing pages for training the models. 

\subsubsection{Limitations} The dependence on webpage screenshots in the inference phase is a hurdle for the deployment of screenshot-based phishing detection solutions. 
In order to take a screenshot, the webpage has to be first fully loaded on a browser. This end-to-end process can take around 2~seconds, which is expensive for a real-time prediction. 
Finally, object detection models are prone to adversarial attacks, e.g., gradient-based attacks such as DeepFool. Phishing attackers may use logos that look visually similar to those of protected brands, but yet defeat the underlying object recognition model used in current phishing solutions. Countering adversarial attacks is an active area of research, and simple techniques, such as changing the activation to step ReLU, are effective against some well-known attacks~\cite{phishpedia-usenix-sec-2021}. Practitioners need to be aware of emerging adversarial attacks to continuously enhance deployed models for  effective mitigation.
     
\subsection{Comparison of different phishing solutions}

Table~\ref{tab:taxonomy} presents the key advantages and disadvantages of the different phishing detection solutions.
While each approach has been shown to have high accuracy on the dataset considered for the corresponding study (e.g., URLNet~\cite{URLnet-2018}, CANTINA+~\cite{canita+-2011}, and Phishpedia~\cite{phishpedia-usenix-sec-2021}), few works have compared across the different methods. 
Yet, intuitively it is clear that content-based solutions outperform URL-based ones, simply because the latter consists of only a small subset of useful features (or information) used by the former.
Evaluations in~\cite{stack-model-phishing-2019} show that the HTML features contribute significantly to the high detection accuracy (Recall) of $\approx97\%$ at an FPR of $\approx0.016$ achieved by the model (a similar observation is made in~\cite{phishing-NDSS-MADWeb-2020}). 
These studies were performed on offline datasets, as is often the case. But that also means, the biases, errors, obsolete links due to the short lifespan of phishing webpages, etc. hardly affect the performance of the models being studied, as they are trained and tested on (different partitions of) the same dataset. In practice, such supervised models need to be regularly retrained to be effective.

Recent work conducted an experiment for one month on the Internet, to evaluate the capabilities of different phishing detectors~\cite[Section~6]{phishpedia-usenix-sec-2021}. Of the top 1000-predictions, the content-based (using both URL and HTML features) had only~9 True Positives; whereas, the screenshot-based phishing detector had more than~900 True Positives. While the study limited validation to the top-1000 predictions, the performance gap between the two models is also exacerbated by the fact that the content-based model is trained only once, in an  offline manner, using a labeled dataset which is usually noisy. 
Besides, since phishing attacks evolve rapidly, the performance of such models degrades quickly with time. On the other hand, the key advantage of the screenshot-based model Phishpedia~\cite{phishpedia-usenix-sec-2021} is that, it  learns the logos of brands from a small reference list ($< 200$), and, importantly, does not depend on a labeled phishing dataset; this explains its high detection capability (also confirmed in another work~\cite{phishintention-usenix-sec-2022}). Based on the above studies, note that the URL-based and content-based models need to be retrained frequently using fresh datasets to keep the model up-to-date. Google trains its content-based classifier  daily~\cite{Google-content-phishing-2010}; and generally in industry, the URL-based and content-based models are retrained at least once a week, using millions of new URLs, so as to mitigate the problem of performance degradation of models. 

\section{Deploying Phishing Detection Models}

\subsection{Using multiple solutions to complement each other}

While URL-based phishing detection is the fastest, content-based and screenshot-based solutions learn from more informative data (page content and screenshots, respectively). But the latter two incur latency, as a webpage has to be fully loaded before the models can be applied. The screenshot-based model has to additionally capture the screenshot, and subsequently perform operations (e.g., logo detection and identification in Phishpedia~\cite{phishpedia-usenix-sec-2021}) on the screenshot before making a prediction. 
One technique to take advantage of these models is to build a pipeline of all three models (working on three different data sources).

Consider a phishing detection pipeline, wherein URLs are first fed to a URL-based solution (e.g., URLNet~\cite{URLnet-2018}) in real-time. In the next stage, the phishing URLs with high prediction probability (say, $\ge \theta$) are streamed to a content-based model (e.g., CANTINA+~\cite{canita+-2011}). Thus, the number of URLs for which entire pages will be fetched is controlled using the threshold $\theta$. 
Since a content-based detector has many more informative features, it can effectively reduce FPs due to URL-based model. 
Yet, as a content-based model is weak when dealing with dynamic content, a screenshot-based solution can come next in the pipeline. For pages for which a content-based solution is weak (e.g., rules can detect the presence of scripts in HTML code), the contents can be rendered on a browser,  screenshot taken, to stream to a screenshot-based solution (e.g., Phishpedia~\cite{phishpedia-usenix-sec-2021}). Such a pipeline leverages the high detection capability (low FPR) of a screenshot-based solution, while still putting it to use only for the most difficult pages to classify. The throughput of this pipeline can be controlled by deciding what fraction of the URLs/pages should be streamed to the next stage.

Finally, in practice, security solutions are often multi-layered. To minimize FPs  and increase the detection rate, classification models are complemented with lists of known benign websites, scoring based on different threat intelligence sources (where we see increasing collaborations within the cyber security industry), and the reputation of IP addresses, domains, and hosting services. Next, we discuss specific deployment options.

\subsection{Integration with Secure Email Gateways}
One of the most common ways of delivering phishing URLs to users is via emails. Enterprises today deploy secure email gateways that analyze the emails being delivered to their employees, to detect, warn and block spam, malicious attachments, and phishing emails. 
Since email is an asynchronous application that tolerates latency, phishing detection solutions based on URL, HTML content, and screenshots, are all good candidates for deployment. Besides, two or more models (along with rules and scoring systems) can be deployed as a pipeline.

Besides URLs, emails contain other important components that are indicative of phishing attacks---headers and subject,  textual content, and attachments. 
These attack surfaces are exploited by attackers for different purposes.
Automatically generating and sending phishing emails in large numbers leave some noticeable attack imprints on the email headers and subjects. But, if an attacker's goal is to lure a particular user into replying with confidential information, then the email body would be exploited with specific contents to persuade the user. A solution that extracts information from multiple components of emails has the potential to detect different kinds of phishing attacks~\cite{d-fence-2021} and is naturally aligned to be integrated with secure email gateways for  detecting phishing emails.

\subsection{Integration with Browsers} 

For browsing users, phishing inference should be made before a webpage is loaded in a browser.
Users typically expect the webpages to load within a few hundred milliseconds.  Thus, screenshot-based phishing detection is not a good candidate, since capturing the screenshot itself requires the webpage to be fully loaded; and this usually takes a couple of seconds, due to network latency, multiple components of the page being loaded from different servers, JavaScript execution, displaying of complex objects, etc. Between URL-based and content-based solutions, the former is better suited for integration with browsers, as they process only URLs to make an inference. A content-based model can still be put to use---Google deploys a content-based model to classify URLs that it receives via other sources (such as Gmail) to build a blacklist~\cite{Google-content-phishing-2010}, and this blacklist is integrated with Google's Chrome browser.

\subsection{Independent Analyzers}
All phishing solutions can run as independent analyzers processing URLs delivered to them. VirusTotal (www.virustotal.com)  
runs tens of engines contributed by different vendors, to analyze user-submitted URLs (which is rate-limited without paid service subscription). Integrating with such a service, a phishing detection solution would get new and potentially suspicious URLs submitted by users the world over. 
The latency requirement here is not as stringent as for browsers, since users are typically prepared to wait for a few seconds for the results. Thus, URL-based and content-based phishing detectors are good candidates for integration with systems like VirusTotal. Alternately, as demonstrated in~\cite{phishpedia-usenix-sec-2021,phishintention-usenix-sec-2022}, solutions can integrate with CertStream (https://certstream.calidog.io/), a service that provides feed of URLs with newly issued certificates in real-time.

\section{Conclusions}

We discussed different models for detecting phishing attacks, based on the input considered---URLs, HTML content, website screenshots, and emails. We also presented practical deployment scenarios for the models. Based on the current state of research as well as experiences from the industry, we conclude with the following takeaways: 
1)~By automatically learning from large-scale data, learning models form an important set of tools to counter phishing attacks. 2)~Phishing detection models have to achieve good Recall at very low FPR for usable security; from our experience, an FPR of ${10}^{-3}$ is usually expected for adoption in industry. 3)~Besides being evaluated on offline datasets, a longitudinal study in the wild is useful for ensuring the proposed model can be generalized. 4)~Different approaches (discussed here) have different pros and cons; besides, learning models are inherently probabilistic and not perfect. Therefore, a pipeline of multi-layered approaches, including rules and scores based on other conventional sources, is required in practice. 
5)~Phishing detection models are prone to different evasion strategies that are technically feasible; in fact, we have evidence of phishing attacks with smart evasion techniques~\cite{trustwave-phishing-campaign-2021}. We identify two important hurdles for phishing detection solutions: i)~cloaking techniques employed by attackers~\cite{zhang2021crawlphish}, ii)~adversarial attacks that target Computer Vision-based deep learning models. Research into next-generation anti-phishing systems should focus on robust approaches for overcoming these challenges while accounting for the ever-evolving nature of phishing content itself.

\bibliographystyle{ieeetr}
\bibliography{main.bib}

\begin{IEEEbiography}{Dinil Mon Divakaran}
(Senior Member, IEEE; dinil.divakaran@trustwave.com) is a senior security researcher at Trustwave, heading R\&D within the CTO office. He is also an Adjunct Assistant Professor of the School of Computing, National University of Singapore (NUS). 
He has more than a decade of experience leading research teams in both industry and academia, working on topics such as phishing detection, network anomaly detection, large-scale security log analysis, DNS security and privacy, IoT network security, programmable in-network security, etc. 
He carried out his doctoral studies at ENS Lyon, France, in the joint lab of INRIA and Bell Labs. His research interests revolve around the broad areas of network and system security, AI for security, protocol analysis, and programmable data planes. 
\end{IEEEbiography}

\begin{IEEEbiography}{Adam Oest} (aoest@paypal.com)
is a senior cybersecurity researcher at PayPal, Inc., where he leads a team focused on the identification and mitigation of sophisticated fraud while simultaneously strengthening the security of the broader web ecosystem. He received his Ph.D. in Computer Science from Arizona State University. His recent work on anti-phishing has won Distinguished Paper Awards at \textit{USENIX Security} and \textit{IEEE Security \& Privacy}, and he was lead author of 2020 Internet Defense Prize winning paper, "Sunrise to Sunset: Analyzing the End-to-end Life Cycle and Effectiveness of Phishing Attacks at Scale," which presents a framework for the proactive detection and end-to-end measurement of phishing attacks.
He is also a Principal Investigator for the PhishFarm Block List Latency Monitoring Program with the Anti-Phishing Working Group (APWG). 
\end{IEEEbiography}

\end{document}